\documentclass[
 twocolumn,
 superscriptaddress,
 amsmath,amssymb,
 aps,
notesatfoot
]{revtex4-2}

\usepackage{graphicx}
\usepackage{dcolumn}
\usepackage{bm}
\usepackage{subcaption}
\usepackage{ragged2e}  

\begin{document}

\preprint{APS/123-QED}



\title{Improving the loss threshold for quantum advantage in photonic sensors by complete photon counting}



\author{Gerard J. Machado}\email{gerard.jimenezmachado@physics.ox.ac.uk}
\affiliation{Clarendon Laboratory, University of Oxford, Parks Road, Oxford OX1 3PU, UK}
\affiliation{Department of Physics, Imperial College London, Prince Consort Road, London, SW7 2AZ, UK}

\author{Yazeed K. Alwehaibi}
\affiliation{Department of Physics, Imperial College London, Prince Consort Road, London, SW7 2AZ, UK}

\author{Guillaume Thekkadath}
\affiliation{National Research Council of Canada, 100 Sussex Drive, Ottawa, Ontario K1A 0R6, Canada}

\author{Zhenghao Li}
\affiliation{Department of Physics, Imperial College London, Prince Consort Road, London, SW7 2AZ, UK}

\author{Aonan Zhang}
\affiliation{Clarendon Laboratory, University of Oxford, Parks Road, Oxford OX1 3PU, UK}

\author{Shang Yu}
\affiliation{Department of Physics, Imperial College London, Prince Consort Road, London, SW7 2AZ, UK}


\author{Adriana E. Lita}
\affiliation{National Institute of Standards and Technology, Boulder, CO, USA}

\author{Richard P. Mirin}
\affiliation{National Institute of Standards and Technology, Boulder, CO, USA}

\author{Martin J. Stevens}
\affiliation{National Institute of Standards and Technology, Boulder, CO, USA}

\author{Juan P. Torres}
\affiliation{ICFO - Institut de Ciencies Fotoniques, The Barcelona Institute of Science and Technology, 08860 Castelldefels (Barcelona), Spain}
\affiliation{Departament of Signal Theory and Communications, Universitat Politecnica de Catalunya, 08034 Barcelona, Spain}

\author{Raj B. Patel}
\affiliation{Department of Physics, Imperial College London, Prince Consort Road, London, SW7 2AZ, UK}

\author{Ian A. Walmsley}
\affiliation{Clarendon Laboratory, University of Oxford, Parks Road, Oxford OX1 3PU, UK}
\affiliation{Department of Physics, Imperial College London, Prince Consort Road, London, SW7 2AZ, UK}

\date{\today}

\begin{abstract}


Tolerance to imperfections is a defining performance criterion for quantum sensors. The threshold for achieving a quantum advantage depends on the input state, sensor configuration, detection scheme, and, critically for optical platforms, photon loss. We consider a nonlinear interferometer in which two gain-optimized parametric nonlinear optical processes couple the state to the internal sensor and subsequently mix the reference and sensor beams. We demonstrate that measuring the full photon-number output statistics of this setup yields marked improvements in the loss threshold. Using photon-number-resolving detection (PNRD) based on transition-edge sensors (TESs), we experimentally reconstruct the joint photon-number statistics at the interferometer output. Subject to internal and external losses of approximately 25\% and 45\%, respectively---and without any post-selection or loss correction---we observe an unconditional violation of the shot-noise limit by $2.37 \pm 0.11$~dB. This translates to a 44\% enhancement in estimation precision over conventional click-detection strategies. We verify this performance by evaluating the classical Fisher information against both an analytical model of the joint photon-number distribution and the raw measured statistics. Ultimately, our results demonstrate that combining nonlinear interferometry with PNRD unlocks metrological information fundamentally inaccessible to click detectors, establishing a clear path toward practical, quantum-enhanced sensing under realistic loss conditions.

\end{abstract}

\maketitle





\section{Introduction}
Optical interferometry can be used to estimate the phase difference $\Delta \varphi$ accumulated between two optical paths, enabling precision measurements of physical quantities ranging from gravitational waves~\cite{aasi2013, tse2019} to the properties of biological tissue~\cite{taylor2013,taylor2016}. In applications where the photon budget is critical, for example, when probing fragile or photosensitive samples~\cite{laissue2017}, minimizing probe power is essential. This makes sensitivity per photon, rather than total sensitivity, the relevant figure of merit. In conventional interferometers utilizing coherent light, the sensitivity of phase estimation is bounded by the shot-noise limit (SNL), given by $\sigma_{\varphi}= 1/\sqrt{N}$, where $\sigma_{\varphi}$ is the phase-estimation uncertainty and $N$ is the mean photon number. Non-classical states of light, such as squeezed and entangled states, can surpass this bound by up to an additional factor of $1/\sqrt{N}$, yielding $\sigma_{\varphi}=1/N$, known as the Heisenberg limit~\cite{caves1981,giovannetti2004,giovannetti2011,demkowicz2015}.

Although quantum states of light can provide enhanced phase sensitivity, in contrast to classical light they are generally non-resilient under \textit{realistic} experimental conditions. The \textit{potential} quantum advantage achievable under ideal conditions is partially lost after considering photon loss, imperfect state preparation, detection inefficiencies, and many other experimental imperfections~\cite{resch2007,thomaspeter2011, slussarenko2017}. In many setups, the advantage provided by quantum correlations is completely erased by modest losses of just a few percent~\cite{demkowicz2013}. Achieving an \textit{unconditional} quantum advantage---one that strictly outperforms classical bounds without post-selecting data or artificially subtracting background losses---is notoriously demanding in photonic metrology. 

Several approaches have been developed to improve loss tolerance, including using adaptive feedback, preparing non-maximally entangled states that are more robust to loss, and engineering low-loss or loss-compensated sensors. These efforts have shown that a quantum enhancement of sensitivity can be achieved with careful attention to both the input state and loss management. What has received less attention in achieving a practical advantage is the measurement strategy tailored to the sensor design. 

In this paper, we demonstrate that sensing beyond the SNL is improved by using a multiphoton measurement strategy, even in a sensor with significant loss. We do this by means of a nonlinear interferometric sensor, which partially compensates loss through coherent amplification, combined with photon-number resolving (PNR) detection. This measurement strategy retains phase information distributed across multiple photon-number components, each contributing to the final phase estimate.

Of course, a lossy sensor will always operate below the Heisenberg limit, even with suitably engineered probe states. For instance, while N00N states---entangled superpositions of zero and $N$ photons---can ideally saturate the Heisenberg limit, they lose all metrological advantage upon the loss of a single photon~\cite{dowling2008}. Holland--Burnett and related photon-number states trade ideal sensitivity for greater robustness to loss~\cite{holland1993,dorner2009,thekkadath2020}. Squeezed-light is also loss-tolerant to a degree and its application in interferometry has enabled practical quantum enhancement in large-scale sensors~\cite{ligo2011,tse2019}. 

Nonlinear interferometers provide an emerging route to loss-resilient quantum sensing. In a so-called $\text{SU}(1,1)$ interferometer~\cite{yurke1986}, the beam splitters of a conventional interferometer are replaced by optical parametric amplifiers. The first nonlinear interaction generates correlated signal and idler fields, while the second one recombines them through phase-sensitive amplification. This architecture can enhance phase sensitivity and reduce the impact of optical attenuation within the system~\cite{chekhova2016,marino2012,ou2012,hudelist2014}. Crucially, an asymmetric configuration of parametric gains offers tailored resilience depending on the dominant loss mechanism~\cite{giese2017}. While a stronger first amplifier mitigates internal sample losses, operating the second amplifier at a higher gain than the first ($G_2 > G_1$) considerably extends tolerance against external detection inefficiencies~\cite{manceau2017,frascella2021, kranias2025, giese2026}. This strategy enables sub-shot-noise sensitivity even under severe detection constraints, making it highly effective when non-ideal readouts present the primary metrological bottleneck.

The remaining sensor improvement possibility lies in the readout. The output of a nonlinear interferometer contains phase information distributed across joint multiphoton correlations. A detection strategy that utilises this information fully can provide additional estimation robustness against loss. While threshold click-detectors can deliver quantum advantages under nearly ideal, loss-free conditions~\cite{qin2023}, they necessarily discard this rich photon-number statistics into a binary response~\cite{heilmann2016, wildfeuer2009}. This compression discards critical metrological information carried by higher-order Fock states, which becomes inaccessible to any click-based estimator~\cite{you2021}. Photon-number-resolving detection~\cite{lita2008}, in contrast, provides direct access to the joint distribution $p_{mn}$ within the relevant multiphoton subspace. By resolving individual Fock layers up to the detector's saturation limit, PNR detection unlocks phase information that is fundamentally lost in conventional threshold detection. Although PNR detection in a lossy sensor cannot reach the ultimate information limit permitted by ideal quantum mechanics, it achieves a stronger violation of the SNL than click detection. In this way, quantum-advantaged sensing can survive in real-world environments, enabling low-light sensing in the presence of unavoidable attenuation. 

We combine an $\text{SU}(1,1)$ interferometer with PNR detection to demonstrate unconditional quantum advantage in loss-resilient phase estimation. Specifically, we directly measure the joint photon-number distribution at the output of the nonlinear interferometer, thus capturing multiphoton correlations that would otherwise be compressed by threshold devices. We derive a closed-form expression for the output distribution and evaluate the classical Fisher information directly from the observed joint statistics. From the raw measured photon-counting data, we observe an unconditional $2.37 \pm 0.11~\text{dB}$ violation of the shot-noise limit, without post-selection or loss correction, together with a $44\%$ enhancement in Fisher information over conventional click-detection strategies. Repeating the experiment across six parametric gain settings confirms that this advantage grows systematically with gain, from $0\%$ to $44\%$, in quantitative agreement with our analytical model. Our results overcome a long-standing constraint in quantum metrology, opening a clear path towards practical quantum-enhanced sensing under non-ideal, real-world operational constraints.

\section{Concept and experimental implementation}

Our experiment implements an SU(1,1) nonlinear interferometer together with PNR readout using transition-edge sensors (TESs). This enables us to map phase variations onto the joint photon-number distribution $p_{mn}(\varphi)$ at the interferometer output. By preserving the multiphoton correlations  that carry the metrological advantage, this approach recovers the information that is lost in conventional binary detection.

\begin{figure*}[t]
\centering
\includegraphics[width=0.70\textwidth]{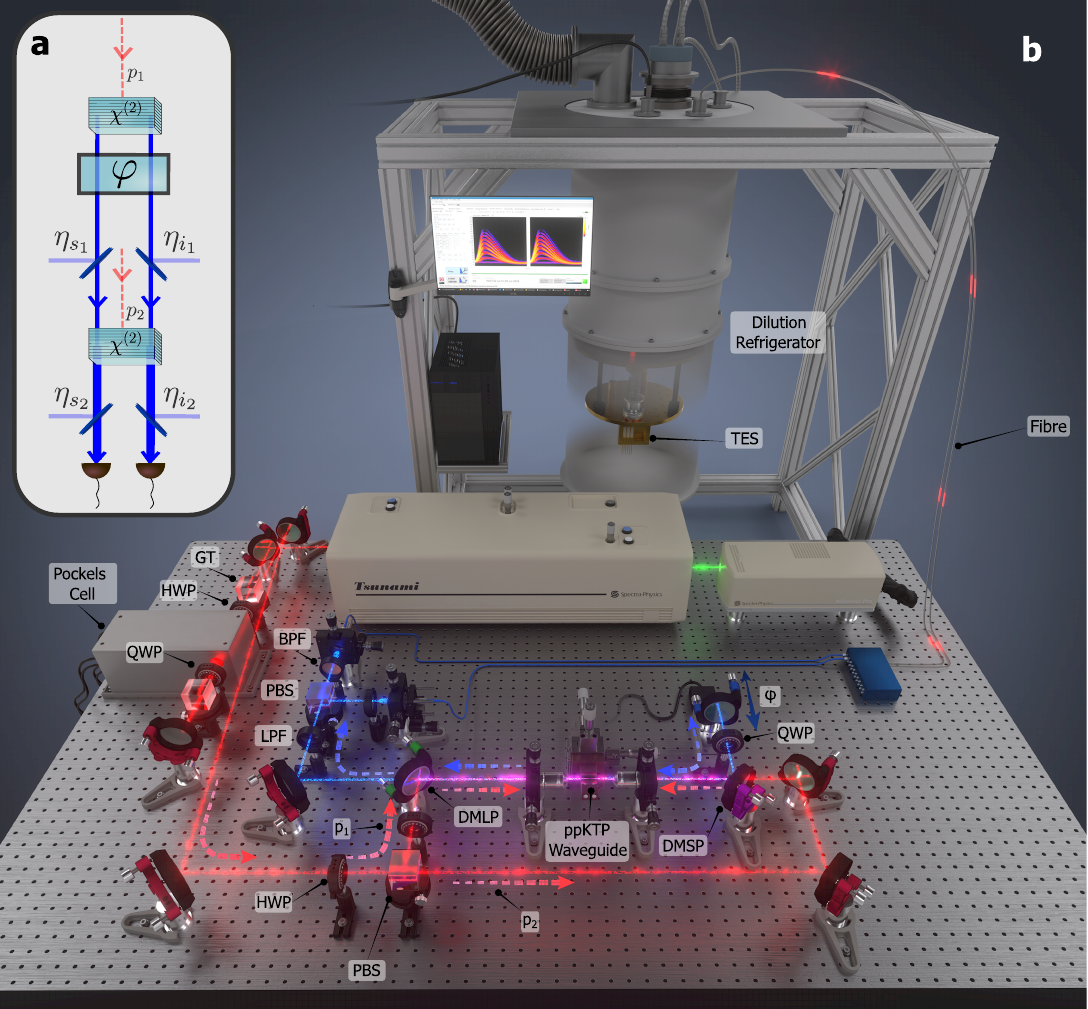}
\caption{\justifying \textbf{Conceptual framework and experimental setup.} \textbf{(a)} Operating principle of the $\text{SU}(1,1)$ nonlinear interferometer. Correlated signal and idler modes from the first parametric stage (gain $G_{1}$) acquire a phase shift $\varphi$ subject to internal transmissions $\eta_{s_1}$ and $\eta_{i_1}$ before seeding a second stage (gain $G_{2}$). External efficiencies $\eta_{s_2}$ and $\eta_{i_2}$ account for optical propagation, fibre coupling, and detector efficiency. \textbf{(b)} Detailed experimental implementation. Pump pulses ($5~\text{ps}$, $80~\text{MHz}$) from a mode-locked Ti:Sapphire laser are pulse-picked to $100~\text{kHz}$ via a double-pass Pockels cell between half-wave plates (HWPs) and orthogonal Glan--Taylor polarisers (GTs). The resulting pulses ($779.3 \pm 0.25~\text{nm}$) are split into paths $p_1$ and $p_2$ to drive successive Type-II parametric down-conversion (PDC) interactions within a 10-mm periodically poled KTP (ppKTP) waveguide, generating degenerate modes at $1558.6~\text{nm}$. Between passes, the internal phase $\varphi$ is scanned via a motorized translation stage, and a quarter-wave plate (QWP) swaps the orthogonal polarisations to compensate for birefringent walk-off, ensuring optimal temporal mode-matching. Residual pump light is removed using dichroic mirrors (DMs) and a long-pass filter (LPF). Output modes are separated by a polarizing beam splitter (PBS), spectrally filtered via bandpass filters (BPFs), fibre-coupled, and routed to transition-edge sensors (TESs) inside a $70~\text{mK}$ dilution refrigerator to reconstruct the joint photon-number distribution $p_{mn}(\varphi)$.}
\label{fig:setup}
\end{figure*}

The operational principle of our scheme is illustrated in Fig.~\ref{fig:setup}(a). The first parametric interaction populates the signal and idler modes with highly correlated photon pairs via spontaneous parametric down-conversion (SPDC). After propagating through the sensing region and acquiring an internal relative phase $\varphi$, these modes undergo a second parametric interaction driven by a secondary pump pulse. This second pass closes the effective $\text{SU}(1,1)$ interferometer, embedding the phase information directly into the multiphoton output statistics of the combined states.

The physical implementation of this architecture is detailed in Fig.~\ref{fig:setup}(b). We utilize a single-crystal, double-pass geometry based on a $10\text{-mm-long}$ periodically poled KTP waveguide. The device is engineered for Type-II phase matching, yielding degenerate signal and idler photons centred at $1558.6~\text{nm}$ when driven by $5~\text{ps}$ pump pulses at $779.3~\text{nm}$. After the first pass, instead of recycling the residual pump—which undergoes significant spatial profile degradation due to the multimode nature of the waveguide at the pump wavelength—we implement two independent pump arms, $p_1$ and $p_2$. This dual-pump configuration ensures optimal spatial mode matching for both interactions. Crucially, it also grants independent control over the respective parametric gains, providing the experimental freedom required to operate in the gain-unbalanced regime that maximizes loss resilience.

Control of the interferometric phase $\varphi$ is achieved by shifting a mirror mounted on a motorized translation stage positioned between the two passes, which modifies the relative phase difference between the down-converted modes and the pump field driving the second parametric amplification stage. Following this second stage, residual pump light is filtered out using a combination of dichroic mirrors and long-pass filters. The co-propagating, orthogonally polarized signal and idler modes are subsequently separated by a polarizing beam splitter, spectrally filtered with bandpass filters to suppress the sinc-sidelobes of the parametric process while preserving high system transmission, and finally injected into single-mode fibres.

The output state is characterized via concurrent PNR detection using two transition-edge sensors (TESs) operating in a dilution refrigerator at $70~\text{mK}$ (see Appendix~\ref{app:tes} for more details). For each experimental shot, the system registers a discrete photon-number pair $(m,n)$. By accumulating these events, we directly reconstruct the joint photon-number distribution $p_{mn}(\varphi)$ up to $m,n = 7$. In stark contrast to threshold click-detectors, which collapse this rich multi-dimensional space into a coarse $2 \times 2$ binary outcome, our PNR approach preserves the high-order multiphoton correlations strictly necessary to extract the full metrological advantage.

\section{Analytical Model of the Output photon Statistics}
Beyond its application to our specific experimental platform, establishing an exact analytical description of photon statistics in  lossy, gain-unbalanced nonlinear interferometers has remained a long-standing challenge due to the complex interplay between attenuation and multiphoton generation. 

To address this issue, we develop an analytical framework that enables the transition from the Heisenberg picture---describing the quantum evolution of operators in the presence of loss---to the Schrödinger-picture allowing state projection onto the Fock basis. By mapping the full quantum dynamics onto a bivariate characteristic function~\cite{glauber1967a, glauber1967b}, we derive a closed-form analytical expression for the joint photon-number distribution $p_{mn}(\varphi)$ that completely bypasses the need for numerical simulations in a truncated Hilbert-space. This framework not only provides the exact tool required to interpret our experimental data and evaluate the classical Fisher information, but also offers a generalized methodology applicable to broader configurations of lossy Gaussian states.

The first parametric interaction (SPDC in this instance) can be described by the Bogoliubov transformations 
\begin{align}
\hat{a}_{s_1} &= U_{s_1} \hat{b}_{s} + V_{s_1} \hat{b}_{i}^{\dagger}, \nonumber \\
\hat{a}_{i_1} &= U_{i_1} \hat{b}_{i} + V_{i_1} \hat{b}_{s}^{\dagger}.
\label{eq:bogo1}
\end{align}

\noindent Here, $\hat{b}_s$ and $\hat{b}_i$ denote the input operators associated with the signal and idler modes, while $\hat{a}_{s_1}$ and $\hat{a}_{i_1}$ are the corresponding output operators. The transformation amplitudes are $U_{s_1,i_1} = \cosh(G_1) \exp(i k_{s,i} L)$ and $V_{s_1,i_1} = -i \sinh(G_1) \exp(i k_{s,i} L + i \varphi_p)$, where $G_1$ is the parametric gain (or squeezing parameter) of the first nonlinear interaction, $L$ is the nonlinear crystal length, $k_{s,i}$ are the wavenumbers of the signal and idler photons, and $\varphi_p$ is the phase of the pump beam. 

To construct a realistic model, we account for both optical attenuation between the two parametric interactions (internal transmissions $\eta_{s_1}$ and $\eta_{i_1}$) and external transmission efficiencies at the detection stage ($\eta_{s_2}$ and $\eta_{i_2}$). As the signal and idler fields propagate in free space, they acquire an unknown phase shift before being reflected back into the nonlinear crystal. The second parametric interaction (parametric amplification) is described by transformations similar to Eq.~\eqref{eq:bogo1}. 

The output quantum operators $\hat{a}_{s_2}$ and $\hat{a}_{i_2}$, associated with the signal and idler modes after the second parametric interaction, can be expressed as
\begin{align}
\hat{a}_{s_2} &= A_{s} \hat{b}_{s} + B_{s} \hat{f}_{s} + C_{s} \hat{b}_{i}^{\dagger} + D_{s} \hat{f}_{i}^{\dagger} + \hat{g}_{s}, \nonumber \\
\hat{a}_{i_2} &= A_{i} \hat{b}_{i} + B_{i} \hat{f}_{i} + C_{i} \hat{b}_{s}^{\dagger} + D_{i} \hat{f}_{s}^{\dagger} + \hat{g}_{i},
\label{eq:output_fields}
\end{align}
where $\hat{f}_{s,i}$ are the quantum noise operators associated with the internal losses~\cite{haus2000, boyd2008}, and $\hat{g}_{s,i}$ correspond to the detection losses. 

The complex functions $A_{s,i}$, $B_{s,i}$, $C_{s,i}$, and $D_{s,i}$, defined explicitly in Appendix~\ref{app:pmn}, depend on several parameters: the unknown phase shifts $\varphi_s$ and $\varphi_i$ acquired by the signal and idler photons (which we aim to estimate); the phases $\varphi_{p_1}$ and $\varphi_{p_2}$ of the pump beams seeding the two nonlinear processes; and the internal ($\eta_{s_1}$, $\eta_{i_1}$) and external ($\eta_{s_2}$, $\eta_{i_2}$) transmission efficiencies.

Using the framework described in Refs. ~\cite{glauber1967a,glauber1967b}, and Eq.~\eqref{eq:output_fields}, we can derive the analytical expression for the joint photon-number distribution $p_{mn}(\varphi)$:
\begin{align}
p_{mn} &= \frac{1}{\mu_{s}\mu_{i}} \sum_{k=0}^{\infty} \left( \frac{|\gamma|^2}{4\mu_{s}\mu_{i}} \right)^k F\left(-m, k+1; 1; \frac{1}{\mu_{s}}\right) \nonumber \\
&\quad \times F\left(-n, k+1; 1; \frac{1}{\mu_{i}}\right),
\label{eq:joint_distribution}
\end{align}
where $F$ is the hypergeometric function~\cite{gradshteyn2007}; $\mu_{s} = 1 + \langle \hat{N}_{s_2} \rangle$ and $\mu_{i} = 1 + \langle \hat{N}_{i_2} \rangle$, with $\langle \hat{N}_{s_2} \rangle$ and $\langle \hat{N}_{i_2} \rangle$ denoting the mean numbers of output signal and idler photons; and $\gamma = 2 \left[A_s C_i+ \left(1-|\eta_{s_1}|^2 \right)B_s D_i \right]$. Equation~\eqref{eq:joint_distribution} establishes a direct analytical bridge between the experimental parameters and the measured Fock-state statistics, providing the mathematical foundation needed to quantify the extracted classical Fisher information. A detailed derivation is provided in Appendix~\ref{app:pmn}.

\section{Unconditional Quantum-Enhanced Phase Sensing}

The multiphoton landscape of the nonlinear interferometer is mapped by experimentally reconstructing the joint photon-number distribution $p_{mn}(\varphi)$ as a function of the internal phase $\varphi$. This measurement, developed in tandem with our analytical model, establishes a self-consistent framework to resolve the full quantum dynamics of the device. Each experimental shot yields a discrete pair of photon-number outcomes $(m,n)$, corresponding to the simultaneously detected signal and idler Fock states.

\begin{figure*}[t]
\centering
\includegraphics[width=0.99\textwidth]{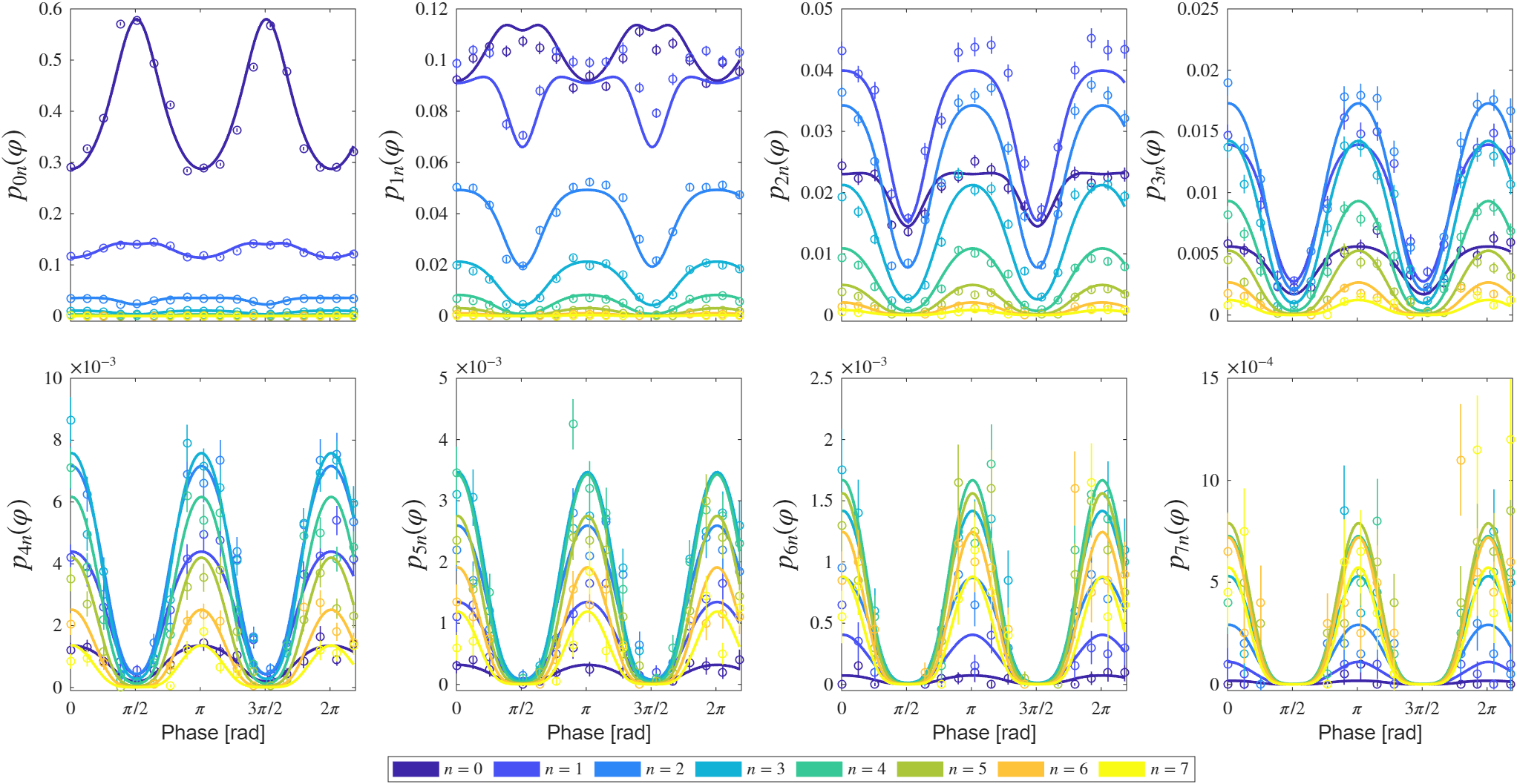}
\caption{\justifying \textbf{Experimental reconstruction of the joint photon-number statistics.} Measured probabilities $p_{mn}(\varphi)$ as a function of the internal phase $\varphi$ for photon-number outcomes up to $m,n=7$. Solid curves represent a single global fit using the analytical model of Eq.~\eqref{eq:joint_distribution}. The phase-dependent multiphoton structure is accurately captured across all dimensions, enabling the self-consistent extraction of the device's internal parameters. Error bars denote one standard deviation, accounting for Poissonian counting statistics and photon-number-assignment uncertainties arising from overlapping TES response distributions.}
\label{fig:joint_statistics}
\end{figure*}

The experimentally reconstructed probabilities for photon-number outcomes up to $m,n=7$ are displayed in Fig.~\ref{fig:joint_statistics}. Crucially, a single global fit based on Eq.~\eqref{eq:joint_distribution} simultaneously captures the phase dependence of the entire multiphoton landscape. The excellent agreement between the experimental data and the theoretical curves demonstrates that our model accurately accounts for the complex interplay of localized parametric gains, internal attenuation, and downstream detection efficiencies.

This global fit yields the physical parameters of our device: the parametric gains $G_1 = 0.499 \pm 0.008$ and $G_2 = 1.044 \pm 0.006$. The transmission coefficients $\eta_{s_1} = 0.758$, $\eta_{i_1} = 0.786$, $\eta_{s_2} = 0.540$, and $\eta_{i_2} = 0.579$ were determined independently from single-pass calibration measurements (see Appendix~\ref{app:calib}). These parameters provide a self-consistent characterization of the device and serve as the foundation for the subsequent Fisher information analysis.

The phase information embedded within the reconstructed multiphoton statistics is rigorously quantified through the classical Fisher information (CFI). For a joint photon-number distribution, the CFI is defined as
\begin{equation}
F(\varphi) = \sum_{m,n} \frac{1}{p_{mn}(\varphi)} \left[ \frac{\partial p_{mn}(\varphi)}{\partial \varphi} \right]^2,
\label{eq:cfi}
\end{equation}
where the sum runs over all accessible signal and idler photon-number sectors. In our analysis, the probabilities $p_{mn}(\varphi)$ are taken directly from the measured statistics, while the phase derivatives are evaluated analytically from our validated model. This avoids the numerical noise amplification inherent in finite-difference differentiation of experimental data, ensuring a robust and conservative CFI estimation.

To establish a rigorous metrological benchmark, we compare our device against the SNL. The SNL is defined as the maximum Fisher information attainable using classical coherent light with the same total mean number of photons actively probing the phase shift inside the interferometer. For our configuration, the total number of photons inside the interferometer is determined by the first parametric interaction, i.e., $\langle \hat{N}_{s_1} \rangle + \langle \hat{N}_{i_1} \rangle$. Therefore, the CFI corresponding to the SNL (see Appendix~\ref{app:cfi}) is:
\begin{equation}
F_{\mathrm{SNL}} = 2\sinh^2(G_1).
\end{equation}
Importantly, this classical baseline is evaluated upstream of any detection losses, meaning that any observed violation represents an absolute, system-level quantum advantage.

\begin{figure*}[t!]
\centering
\includegraphics[width=0.99\textwidth]{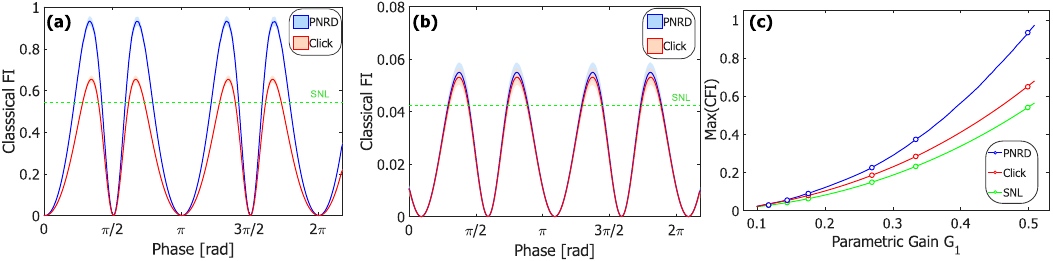}
\caption{\justifying \textbf{Classical Fisher information for phase sensing across parametric gain regimes.} \textbf{(a)}~CFI as a function of interferometric phase at $G_1 = 0.499 \pm 0.008$ and $G_2 = 1.044 \pm 0.006$. The PNR measurement (blue) surpasses the shot-noise limit (green dashed) by $2.37 \pm 0.11$~dB and outperforms click detection (red) by 44\% at the optimal phase, both without loss correction or post-selection. \textbf{(b)}~Same measurement at $G_1 = 0.145 \pm 0.006$ and $G_2 = 0.463 \pm 0.006$, illustrating the weak-gain regime where PNR and click-detection CFI are nearly identical and the quantum advantage over the SNL is marginal. \textbf{(c)}~Maximum CFI as a function of parametric gain $G_1$, measured at six pump powers spanning $G_1 \in [0.12, 0.50]$ and $G_2 \in [0.34, 1.04]$. Points show experimental values for PNRD (blue), click detection (red), and the SNL (green); solid curves show the theoretical prediction evaluated at the experimentally determined parameters including background photons. The advantage of PNR over click detection grows monotonically from 0\%  at $G_1 = 0.12$ to $44\%$ at $G_1 = 0.50$, in excellent agreement with theory across the full gain range. Shaded regions in \textbf{(a)} and \textbf{(b)} represent one standard deviation obtained via error propagation of the experimental uncertainties.}
\label{fig:cfi}
\end{figure*}

In Fig.~\ref{fig:cfi}(a), we plot the experimentally determined CFI as a function of the interferometric phase at $G_1 = 0.499 \pm 0.008$ and $G_2 = 1.044 \pm 0.006$. The full joint photon-number statistics surpass the SNL over a broad phase range, reaching a peak enhancement of $2.37 \pm 0.11~\text{dB}$. This constitutes an unconditional demonstration of quantum-enhanced phase sensitivity under realistic operational losses, achieved without resorting to post-processing techniques. Crucially, this robust advantage under modest external efficiencies ($\eta_{s_2}, \eta_{i_2} \sim 55\%$) is enabled by our gain-unbalanced configuration ($G_2 > G_1$), which amplifies the phase-sensitive multiphoton correlations before they reach the detection channel.

To quantify the specific advantage of photon-number resolution, we compare our results against a threshold-detection benchmark. This baseline is constructed by coarse-graining our experimental data into binary ``click'' or ``no-click'' events. Specifically, we marginalize our measured joint statistics onto the binary subspace $\{(0,0),(m>0,0),(0,n>0),(m>0,n>0)\}$, which replicates the response of threshold detectors. At the optimal phase operating point, the full PNR measurement outperforms this click-detection scheme by 44\%, proving that tracking the higher-order multiphoton correlations recovers substantial phase information that is otherwise discarded by binary detection.

To establish the generality of these results, we repeat the experiment at five additional pump powers, varying the parametric gains over the range $G_1 \in [0.12, 0.50]$ and $G_2 \in [0.34, 1.04]$ while maintaining the same gain-unbalanced configuration and transmission coefficients. Figure~\ref{fig:cfi}(c) shows the maximum CFI as a function of $G_1$ for PNR detection, click detection, and the SNL. The experimental values are in excellent agreement with the theoretical predictions evaluated at the experimentally determined parameters, validating our analytical model across a wide range of operating conditions. The advantage of PNR over click detection grows monotonically with $G_1$, scaling from no advantage in the weak-gain regime ($G_1 = 0.12$, $G_2 = 0.34$) to $44\%$ at the highest gain ($G_1 = 0.50$, $G_2 = 1.04$), consistent with the increasing contribution of higher-order photon-number correlations to the Fisher information as the mean photon number grows.

These results demonstrate that photon-number-resolving measurements recover metrological information that is fundamentally inaccessible to threshold detection, establishing a practical route to unconditional quantum-enhanced sensing with nonlinear interferometers.

\section{Discussion}

The core significance of this demonstration lies in the fact that our quantum advantage is obtained unconditionally—under realistic operational losses and strictly without post-selection or loss correction. Nonlinear interferometers are inherently celebrated for their loss-tolerant sensing architectures~\cite{yurke1986, hudelist2014, qin2023, giese2017, frascella2021}. However, as proven by our threshold-detection comparison, conventional binary readout discards a substantial fraction of the phase information carried by higher-order Fock states. This limitation becomes critical in practical sensing scenarios where the probed sample introduces absorption or scattering losses. While state-of-the-art click-detection implementations have achieved impressive quantum enhancement by operating under nearly ideal laboratory efficiencies ($\sim 97.5\%$ internal transmission in Ref.~\cite{qin2023}), their metrological advantage degrades rapidly when faced with such operational constraints. In contrast, PNR readout provides the missing measurement capability required to extract this remaining phase information efficiently, acting as a robust shield against non-ideal transmission. The present experiment therefore significantly extends previous metrological milestones~\cite{slussarenko2017, thekkadath2020} by establishing a practical paradigm where photon-number resolution preserves quantum advantages directly at the detection stage across a wide range of realistic loss regimes.

Looking forward, the physical principles demonstrated in this work are highly compatible with more integrated, scalable, and stable photonic platforms. Transitioning our architecture to compact waveguide geometries or thin-film lithium niobate-on-insulator (TFLNOI) circuits~\cite{escale2020,zhang2017}, combined with fully integrated, low-loss PNR detection interfaces implemented on-chip~\cite{gerrits2011, lita2008}, should enable phase sensing at larger photon numbers while maintaining long-term phase stability. Beyond standalone phase estimation, this generalized approach of utilizing full joint photon-number statistics under loss holds immediate relevance for broader quantum technologies. For instance, in quantum illumination~\cite{lopaeva2013} and quantum imaging~\cite{giovannetti2011}, where the information extracted per probing photon is the ultimate limiting resource and return-channel losses are severe, the ability to resolve multiphoton correlations without post-selection offers a direct pathway toward robust operation in high-noise, high-loss backgrounds. Similarly, the mathematical and experimental framework developed here could inspire scaling strategies for macroscopic quantum metrology, where multiphoton interference resilience is paramount to surviving real-world environmental perturbations.

\begin{acknowledgments}
We thank K. Heshami for the helpful discussions and feedback on the manuscript. This work was funded by the National Research Council of Canada (project QSP 062-2) and partly funded by UK Research and Innovation Future Leaders Fellowship (project MR/W011794/1), UK Research and Innovation Guarantee Postdoctoral Fellowship (project: EP/Y029631/1), and Department of Science, Innovation \& Technology Tactical Fund. JPT acknowledges support from the project NOVISLIGHT (PID2023-149780NB-I00) funded by Ministerio
de Ciencia, Innovación y Universidades (Proyectos de generación de conocimiento 2023). NIST work was funded solely by the United States Government. 

\end{acknowledgments}
\vspace{2em}
This document has not been peer reviewed but has been cleared by NIST for release.

\appendix


\section{Transition Edge Sensor (TES)}
\label{app:tes}
Photon-number-resolving detection is performed using transition-edge sensors optimized for telecom wavelengths. The active element is a tungsten thin film embedded in an optical cavity designed to enhance absorption near \(1550\,\mathrm{nm}\). The sensors are operated inside an Oxford Instruments Triton 200 dilution refrigerator \footnote{The use of trade names is intended to allow the measurements to be appropriately interpreted and does not imply endorsement by the US government, nor does it imply these are necessarily the best available for the purpose used here.} at a base temperature below \(70\,\mathrm{mK}\).

Each TES is voltage biased in the superconducting transition and read out using a SQUID-array amplifier. Absorption of photons increases the TES resistance, reducing the current through the sensor and producing a flux change that is amplified by the SQUID array. The resulting voltage traces are amplified at room temperature, low-pass filtered at 1~MHz, and digitized using a 14-bit waveform digitizer.

To assign a discrete photon-number label to each individual pulse event, the digitized voltage traces are processed using a pulse-filtering protocol. Maintaining a pulse repetition rate below 200~\text{kHz} ensures full thermal relaxation of the tungsten film between consecutive events, yielding well-isolated voltage vectors $\bm{v}_i$. These raw traces are then categorized using a matched-filter~\cite{levine2012}, computing the overlap $\langle \bm{v}_i, \bm{\bar{v}} \rangle$ against a calibrated average reference trace $\bm{\bar{v}}$. The resulting histogram of these overlap values exhibits distinct, well-resolved peaks, allowing for unambiguous photon-number assignment via standard valley-finding techniques~\cite{zhenghao2025}.

\section{Derivation of the joint photon-number distribution probability $p_{mn}$} 
\label{app:pmn}
Here we provide a step-by-step derivation of Eq.~\eqref{eq:joint_distribution} of the main text, the analytical expression of the joint photon-number distribution $p_{mn}$. We make use of the Heisenberg picture to describe the quantum evolution of operators in the SU(1,1) interferometer. We obtain the probability distribution $p_{mn}$ using the framework described by Mollow and Glauber~\cite{glauber1967a, glauber1967b}, which makes use of the bi-variate characteristic function. 

\subsection{Evolution of Quantum Operators in the Heisenberg Picture}
Eqs. (\ref{eq:bogo1}) in the main text describe the relationship between input ($\hat{b}_{s}$ and $\hat{b}_{i}$)  and output ($\hat{a}_{s_1}$ and $\hat{a}_{i_1}$) quantum operators in the first nonlinear parametric interaction (SPDC). The signal and idler photons generated by means of SPDC propagate through the internal sensing region. They experience phase shifts $\varphi_{s}$ (signal) and $\varphi_i$ (idler), and optical attenuation characterized by the transmission amplitudes $\eta_{s_1}$ and $\eta_{i_1}$. The effect of losses on quantum propagation \cite{boyd2008,haus2000} can be described with the help of quantum operators $\hat{f}_s$ and $\hat{f}_i$ so that
\begin{align}
\hat{a}_{s_1}  \Longrightarrow \eta_{s_1} \exp \left( i \varphi_s \right)\,\hat{a}_{s_1} + \hat{f}_s  \nonumber \\
\hat{a}_{i_1}  \Longrightarrow \eta_{i_1} \exp \left( i\varphi_i \right)\,\hat{a}_{i_1} + \hat{f}_i \label{qloss}
\end{align}
where the quantum operators $\hat{f}_s$ and $\hat{f}_i$ fulfill the commutation relationships $\big[ \hat{f}_{k}, \hat{f}_{k}^{\dagger} \big] = 1 - |\eta_{k_1}|^2$~for $k \in \{s,i\}$.

The second parametric interaction is also described by Eqs. (\ref{eq:bogo1}), between input quantum operators $\hat{a}_{s_1}$ and $\hat{a}_{i_1}$, and quantum output operators $\hat{a}_{s_2}$ and $\hat{a}_{i_2}$. Since the second interaction process will have in general a different parametric gain $G_2$, we have to employ new transformation amplitudes $U_{s_2}$, $U_{i_2}$, $V_{s_2}$, and $V_{i_2}$, where we substitute $G_1 \longrightarrow G_2$. 

To capture detection efficiencies and photon collection losses, we introduce a final external attenuation layer modeled by transmission coefficients $\eta_{s_2}$ and $\eta_{i_2}$ and the operator transformations given by Eqs. (\ref{qloss}), now associated with the quantum operators $\hat{g}_s$ and $\hat{g}_i$. These quantum operators fulfill the commutation relationships $\big[ \hat{g}_{k}, \hat{g}_{k}^{\dagger} \big] = 1 - |\eta_{k_2}|^2$ with  $k \in \{s,i\}$.

Collecting all terms, the ultimate output field operators can be mapped linearly back to the initial input and noise operators by Eqs. (\ref{eq:output_fields})
with 
\begin{eqnarray}
& & A_s= \eta_{s_2}\, \left[ \eta_{s_1}\,U_{s_2}\,U_{s_1} \, \exp(i\varphi_s)+\eta_{i_1}^*\, V_{s_2} V_{i_1}^*\, \exp(-i\varphi_i) \right]  \nonumber \\
& & B_s= \eta_{s_2}\,U_{s_2} \nonumber \\
& & C_s= \eta_{s_2}\,\left[ \eta_{s_1}\,U_{s_2} V_{s_1} \, \exp(i\varphi_s)+ \eta_{i_1}^*\,V_{s_2} U_{i_1}^*  \exp(-i \varphi_i)\right] \nonumber \\
& & D_s=\eta_{s_2}\, V_{s_2} \label{AsBsCsDs}
\end{eqnarray}
and
\begin{eqnarray}
& & A_i=\eta_{i_2}\, \left[ \eta_{i_1}\,U_{i_2}\,U_{i_1} \, \exp(i\varphi_i)+ \eta_{s_1}^*\,V_{i_2} V_{s_1}^*\,\exp(-i\varphi_s) \right] \nonumber \\
& & B_i= \eta_{i_2}\, U_{i_2} \nonumber \\
& & C_i= \eta_{i_2}\, \left[ \tau_{i_1}\,U_{i_2} V_{i_1} \,  \exp(i\varphi_i)+ \eta_{s_1}^*\,V_{i_2} U_{s_1}^* \exp(-i \varphi_s)\right] \nonumber \\
& & D_i=\eta_{i_2}\, V_{i_2} \label{AiBiCiDi}
\end{eqnarray}
The total mean photon number detected at the output of the nonlinear interferometer are $\langle \hat{N}_{s_2} \rangle = \langle \hat{a}_{s_2}^{\dagger} \hat{a}_{s_2} \rangle=|C_s|^2 + |D_s|^2 (1 - |\eta_{i_1}|^2)$ and $\langle \hat{N}_{i_2} \rangle = \langle \hat{a}_{i_2}^{\dagger} \hat{a}_{i_2} \rangle=|C_i|^2 + |D_i|^2 (1 - |\eta_{s_1}|^2)$.

\subsection{Constructing the Characteristic Function}
To extract the discrete counting statistics, we need to evaluate the characteristic function, as function of the variables $\eta$ and $\xi$:
\begin{equation}
\chi(\eta, \xi) = \operatorname{Tr} \left[ \hat{\rho}_0 \exp\left( \eta \hat{a}_{s_2}^{\dagger} + \xi \hat{a}_{i_2}^{\dagger} - \eta^* \hat{a}_{s_2} - \xi^* \hat{a}_{i_2} \right) \right], \label{xi}
\end{equation}
where $\hat{\rho}_0 = |0\rangle\langle 0|$ the initial vacuum state across all entry ports. Substituting in Eq. (\ref{xi}) the expressions of the quantum operators $\hat{a}_{s_2}$ and $\hat{a}_{i_2}$ given by Eqs. (\ref{eq:output_fields}), the exponent can be partitioned into a pair of non-commuting operators, i.e., $\hat{\eta} \hat{a}_{s_2}^{\dagger} + \xi \hat{a}_{i_2}^{\dagger} - \eta^* \hat{a}_{s_2} - \xi^* \hat{a}_{i_2} = \hat{F} + \hat{G}$, where $\hat{F}$ clusters creation operators and $\hat{G}$ clusters annihilation operators. Their commutator reduces to a scalar value:
\begin{align}
& [\hat{F}, \hat{G}] = |\eta|^2 \left(1 + 2 \langle \hat{N}_{s_2} \rangle \right) + |\xi|^2 \left(1 + 2 \langle \hat{N}_{i_2} \rangle \right) + \nonumber \\
& -\eta \xi \gamma^* - \eta^* \xi^* \gamma,
\end{align}
where
\begin{align}
&\gamma = 2 \left[ A_s C_i + \left(1 - |\eta_{s_1}|^2\right) B_s D_i \right] = \nonumber \\
&= 2 \left[ A_i C_s + \left(1 - |\eta_{i_1}|^2\right) B_i D_s \right].
\end{align}
Applying the Baker-Campbell-Hausdorff identity $\exp(\hat{F}+\hat{G}) = \exp(-\frac{1}{2}[\hat{F},\hat{G}])\exp(\hat{F})\exp(\hat{G})$ and noting that $\langle 0 | \exp(\hat{F})\exp(\hat{G}) | 0 \rangle = 1$, the joint characteristic function takes an exact bivariate Gaussian form:
\begin{align}
& \chi(\eta, \xi) = \exp \Big\{ -|\eta|^2 \left[ \frac{1}{2} + \langle \hat{N}_{s_2} \rangle \right]  + \nonumber \\
& -|\xi|^2 \left[ \frac{1}{2} + \langle \hat{N}_{i_2} \rangle \right] -\frac{1}{2}\eta\xi \gamma^* - \frac{1}{2} \eta^* \xi^* \gamma \Big\}.
\label{eq:joint_chi}
\end{align}

\subsection{Fock-Space Projection and Analytical Integration}
The joint probability $p_{mn}$ of detecting $m$ signal photons in coincidence with $n$ idler photons in the idler corresponds to the diagonal elements $\langle m, n | \hat{\rho} | m, n \rangle$ of the density matrix $\hat{\rho}$.  We obtain \cite{glauber1967b}
\begin{eqnarray}
& & p_{mn} = \frac{1}{\pi^2} \iint d^2\eta\, d^2\xi\, \chi(\eta, \xi)\, \exp \left\{ -\frac{1}{2}|\eta|^2 - \frac{1}{2}|\xi|^2 \right\}  \nonumber  \\
& & \times \langle m | \exp \left\{ -\eta \hat{a}_{s_2}^{\dagger} \right\} \exp \left\{ \eta^* \hat{a}_{s_2} \right\} | m \rangle  \label{pmn}\\
& & \times \langle n | \exp \left\{ -\xi \hat{a}_{i_2}^{\dagger} \right\} \exp \left\{ \xi^* \hat{a}_{i_2} \right\} | n \rangle. \nonumber
\end{eqnarray}
If we substitute Eq.~\eqref{eq:joint_chi} into Eq. (\ref{pmn}), and we make use of the identity \cite{barnett} $\langle m | \exp \left\{ -\eta \hat{a}^{\dagger} \right\} \exp \left\{\eta^* \hat{a} \right\} | m \rangle = \exp \left\{ -\frac{1}{2}|\eta|^2 \right\} L_m(|\eta|^2)$, we obtain
\begin{align}
& p_{mn} = \frac{1}{\pi^2} \iint d^2\eta \, d^2\xi \, L_m(|\eta|^2) L_n(|\xi|^2) \times \nonumber \\
&\times \exp \left\{ -\mu_s|\eta|^2 - \mu_i|\xi|^2 - \frac{1}{2}\eta\xi\gamma^* - \frac{1}{2}\eta^*\xi^*\gamma \right\},
\label{eq:prob_integral_raw}
\end{align}
where $\mu_s = 1 + \langle \hat{N}_{s_2} \rangle$ and $\mu_i = 1 + \langle \hat{N}_{i_2} \rangle$.

To compute the integral, we switch to polar coordinates, $\eta = |\eta| \exp \left( i\varphi_1 \right)$ and $\xi = |\xi| \exp \left( i\varphi_2 \right)$, and make the transformations $\varphi_+ = \varphi_1 + \varphi_2$ and $\varphi_- = \varphi_1 - \varphi_2$, which introduces a Jacobian factor of $1/2$. The angular integration can be evaluated analytically via the modified Bessel function identity $\int_0^{2\pi} dx \exp(p\cos x + q\sin x) = 2\pi I_0(\sqrt{p^2+q^2})$, yielding:
\begin{align}
& & \frac{1}{2}\,\int_0^{4\pi} d\varphi_+ \int_0^{2\pi} d\varphi_- \, \exp \Big\{ -|\eta||\xi| \gamma_R \cos(\varphi_+) \Big\} \nonumber\\
& & \times \exp \Big\{ -|\eta||\xi| \gamma_I \sin(\varphi_+)  \Big\} = 4\pi^2 I_0\left(|\eta||\xi||\gamma|\right),
\end{align}
where $\gamma_R$ and $\gamma_I$ are the real and imaginary parts of $\gamma$. This reduces Eq.~\eqref{eq:prob_integral_raw} to a purely radial double integral. Transforming the radial coordinates via $x = |\eta|^2$ and $y = |\xi|^2$, Eq. (\ref{eq:prob_integral_raw}) becomes
\begin{align}
& & p_{mn} = \int_0^{\infty} dx\, \int_0^{\infty} dy\, L_m(x) L_n(y) \nonumber \\
& & \times \exp \left\{ -\mu_s x - \mu_i y \right\}\,I_0\left(\sqrt{xy}|\gamma|\right).
\end{align}
We expand the modified Bessel function into its power series, $I_0(z) = \sum_{k=0}^{\infty} (z/2)^{2k}/(k!)^2$, enabling us to separate the integrations over $x$ and $y$:
\begin{align}
& p_{mn} = \sum_{k=0}^{\infty} \frac{|\gamma|^{2k}}{4^k (k!)^2} \left[ \int_0^{\infty}\!dx \, x^k L_m(x) e^{-\mu_sx} \right] \nonumber \\
&\times \left[ \int_0^{\infty}\!dy \, y^k L_n(y) e^{-\mu_iy} \right].
\label{eq:separated_radial}
\end{align}
Finally, we evaluate each integral using the standard Gradshteyn-Ryzhik integration identity for confluent hypergeometric functions~\cite{gradshteyn2007}:
\begin{equation}
\int_0^{\infty} dt  \exp \left( -st \right)\, t^k L_m(t) = \frac{k!}{s^{k+1}} F \left(-m, k+1, 1, \frac{1}{s} \right).
\end{equation}
Substituting this result into Eq.~\eqref{eq:separated_radial} causes the $(k!)^2$ terms in the denominator to cancel out exactly. Factoring out the leading terms, we arrive at the closed-form analytical expression presented in the main text:
\begin{align}
& p_{mn} = \frac{1}{\mu_s\mu_i} \sum_{k=0}^{\infty} \left( \frac{|\gamma|^2}{4\mu_s\mu_i} \right)^k F\left(-m, k+1; 1; \frac{1}{\mu_s}\right)  \nonumber \\
&\times F\left(-n, k+1; 1; \frac{1}{\mu_i}\right).
\end{align}

\section{Classical Fisher information for phase estimation with coherent states in linear optical interferometers}
\label{app:cfi}
Quantum-enhanced phase estimation is defined in comparison to the sensitivity that one can achieve in phase estimation using conventional optical interferometers and coherent light.
In a Mach-Zehnder or Michelson interferometer, the number of photons at each output port ($1$ and $2$) of the interferometer are 
\begin{equation}
N_1 = N_0 \sin^2 \frac{\varphi}{2}, \hspace{0.5cm} N_2=  N_0 \cos^2 \frac{\varphi}{2},
\end{equation}
where $\varphi$ is the phase unbalance between the two arms of the interferometer, and $N_0$ is the number of photons entering the interferometer through one of the input ports. When considering coherent states, one can demonstrate \cite{vandenbos} that the classical Fisher information for estimation of the phase difference $\varphi$ is
\begin{equation}
F_{\text{coh}}^i = \frac{1}{N_i} \left[ \frac{\partial N_i}{\partial \varphi} \right]^2. 
\end{equation}
If we use output port $1$ for phase estimation, the classical Fisher information is $F_{\text{coh}}^1=N_0/\cos^2 (\varphi/2)$, while using output port $2$ we have $F_{\text{coh}}^2=N_0/\sin^2 (\varphi/2)$. It is straightforward to show that in both cases the maximum value is $F_{\text{coh}}^{\text{max}}=N_0$.

In the SU(1,1) interferometer, the number of photons entering the interferometer, signal/idler photons generated in the first nonlinear parametric interaction, is $N=2 \sinh^2 G_1$, where $G_1$ is the parametric gain. In a conventional interferometer, using coherent light with this mean number of photons, the maximum classical Fisher information (Shot Noise limit) would be $F_{SNL}=2 \sinh^2 (G_1)$.  This is our reference to determine when we have phase estimation with quantum-enhanced sensitivity.

\section{Global fitting procedure}
\label{app:fit}
The analytical model of Eq.~\eqref{eq:joint_distribution}, extended to include Poissonian background photons (see Appendix~\ref{app:background}), is fitted globally to all measured distributions $p_{mn}^{\rm obs}(\varphi)$ simultaneously. The fit minimizes the weighted least-squares cost function
\begin{equation}
\chi^2 = \sum_{m,n,\varphi} 
\frac{\left[p_{mn}^{\rm obs}(\varphi) - p_{mn}^{\rm model}(\varphi)\right]^2}
{\sigma_{mn}^2(\varphi)},
\end{equation}
where $\sigma_{mn}(\varphi)$ denotes the combined experimental uncertainty at each phase point, accounting for Poissonian counting statistics and photon-number-assignment errors from overlapping TES response distributions. The minimization is performed using the Levenberg--Marquardt algorithm as implemented in \textsc{Matlab}'s \texttt{lsqnonlin} routine.

The transmission coefficients $\eta_{s_1}$, $\eta_{i_1}$, $\eta_{s_2}$, and $\eta_{i_2}$ are fixed to values determined from independent single-pass calibration measurements (see Appendix~\ref{app:calib}), reducing the number of free parameters to five: the parametric gains $G_1$ and $G_2$, the background photon numbers $\bar{n}_s^{\rm bg}$ and $\bar{n}_i^{\rm bg}$, and the phase offset $\varphi_0$. Starting values and bounds are chosen to reflect physical constraints ($G_{1,2} > 0$, $\bar{n}^{\rm bg} \geq 0$, $|\varphi_0| \leq \pi/4$).

The global fit converges to $G_1 = 0.499 \pm 0.008$, $G_2 = 1.044 \pm 0.006$, $\varphi_0 = -0.028 \pm 0.006$~rad, $\bar{n}_s^{\rm bg} = 0.079$, and $\bar{n}_i^{\rm bg} = 0.101$, with a reduced chi-squared of $\chi^2_\nu = 5.1$. Parameter uncertainties are estimated from the inverse Hessian of the cost function, scaled by $\sqrt{\chi^2_\nu}$ to account for systematic residuals not fully captured by the single-mode model. The elevated $\chi^2_\nu$ is attributed primarily to systematic residuals in the vacuum component $p_{00}(\varphi)$, which contributes the largest fraction of the CFI at the optimal phase; we verify that this discrepancy does not qualitatively affect the reported metrological conclusions.

\section{Poissonian background model}
\label{app:background}
To account for residual background photons arising from stray light and imperfect spectral filtering, the interferometer output distribution is convolved with independent Poissonian noise terms:
\begin{equation}
p_{mn}^{\rm obs}(\varphi) = \sum_{j=0}^{m}\sum_{k=0}^{n} 
p_{m-j,\,n-k}(\varphi)\, 
\frac{e^{-\bar{n}_s^{\rm bg}}\,(\bar{n}_s^{\rm bg})^j}{j!}\,
\frac{e^{-\bar{n}_i^{\rm bg}}\,(\bar{n}_i^{\rm bg})^k}{k!},
\end{equation}
where $p_{mn}(\varphi)$ is the interferometer distribution of 
Eq.~\eqref{eq:joint_distribution} and $\bar{n}_{s,i}^{\rm bg}$ 
are the mean background photon numbers in the signal and idler channels, respectively. The background parameters are treated as free parameters in the global fit and are consistent with stray-light levels reported for fibre-coupled TES detectors at telecom wavelengths~\cite{li2025}.

\section{Single-pass calibration measurements}
\label{app:calib}
The transmission coefficients $\eta_{s_2}$ and $\eta_{i_2}$,  which encode the external detection efficiencies and coupling losses, and $\eta_{s_1}$ and $\eta_{i_1}$, which parameterize the internal losses between the two parametric interactions, 
are determined from independent single-pass measurements performed at the same pump power as the interferometer experiment.

For each channel, the mean detected photon number is measured as a function of the average pump power $P$ under two configurations: with only the first parametric interaction active (pump $p_2$ blocked), and with only the second interaction active (pump $p_1$ blocked). 

The mean photon number scales as
\begin{equation}
\langle \hat{N} \rangle = \eta^2 \sinh^2(\kappa\sqrt{P}),
\end{equation}
where $\kappa$ is a crystal-dependent nonlinear coupling coefficient and $\eta$ is the relevant transmission~\cite{chekhova2006,spasibko2012}. A nonlinear least-squares fit to the power-dependent data yields $\kappa$ and $\eta^2$ for each channel independently.

Since both parametric interactions use the same waveguide, the nonlinear coupling coefficient $\kappa$ is identical for the two passes. This allows the internal transmission coefficients to be extracted from the ratio of the fitted amplitudes:
\begin{equation}
\eta_{s_1} = \sqrt{\frac{\langle\hat{N}_{s,\,p_2\,\rm blocked}\rangle}
{\langle\hat{N}_{s,\,p_1\,\rm blocked}\rangle}}, \quad
\eta_{i_1} = \sqrt{\frac{\langle\hat{N}_{i,\,p_2\,\rm blocked}\rangle}
{\langle\hat{N}_{i,\,p_1\,\rm blocked}\rangle}}.
\end{equation}
The external coefficients $\eta_{s_2}$ and $\eta_{i_2}$ are obtained directly from the fitted amplitudes of the second-pass measurements, using the parametric gain $G_2 = \kappa\sqrt{P_{\rm exp}}$ evaluated at the experimental pump power $P_{\rm exp}$. The calibration yields $\eta_{s_1} = 0.758$, $\eta_{i_1} = 0.786$, $\eta_{s_2} = 0.540$, and $\eta_{i_2} = 0.579$, with fit quality $R^2 > 0.999$ for the second-pass channels and $R^2 > 0.966$ for the first-pass channels. The lower $R^2$ for the first-pass measurements reflects the smaller detected photon numbers and correspondingly larger relative uncertainties at low pump powers.

\nocite{*}

\bibliography{apssamp}

@PREAMBLE{
 "\providecommand{\noopsort}[1]{}" 
 # "\providecommand{\singleletter}[1]{#1}%" 
}

@BOOK{Bire82,
   author       = {N. D. Birell and P. C. W. Davies},
   year         = 1982,
   title        = {Quantum Fields in Curved Space},
   publisher    = {Cambridge University Press}
}

@ARTICLE{feyn54,
   author       = "R. P. Feynman",
   year         = "1954",
   journal      = "Phys.\ Rev.",
   volume       = "94",
   pages        = "262",
   doi          = "10.1029/2002JD002268",
}

@ARTICLE{epr,
   author       = "A. Einstein and {\relax Yu} Podolsky and N. Rosen", 
   year         = "1935", 
   journal      = "Phys.\ Rev.", 
   volume       = "47", 
   pages        = "777",
}

@ARTICLE{Berman1983,
   author       = "Berman, Jr., G. P. and Izrailev, Jr., F. M.",
   title        = "Stability of nonlinear modes",
   journal      = "Physica D",
   volume       = "88", 
   pages        = "445",
   year         = "1983",
}

@ARTICLE{Davies1998,
   author       = "E. B. Davies and L. Parns", 
   title        = "Trapped modes in acoustic waveguides", 
   journal      = "Q. J. Mech. Appl. Math.", 
   volume       = "51", 
   pages        = "477--492", 
   year         = "1988", 
}

@MISC{witten2001,
   author       = "Edward Witten",
   eprint       = "hep-th/0106109",
   year         = "2001", 
}

@INBOOK{Beutler1994,
   author       = "E. Beutler", 
   editor       = "E. Beutler and M. A. Lichtman and B. W. Coller and T. S. Kipps", 
   title        = "Williams Hematology", 
   chapter      = "7", 
   pages        = "654--662",
   publisher    = "McGraw-Hill", 
   year         = "1994", 

   edition      = "5", 
   address      = "New York", 
   volume       = "2", 
}

@INBOOK{inbook-full,
   author = "Donald E. Knuth",
   title = "Fundamental Algorithms",
   volume = 1,
   series = "The Art of Computer Programming",
   publisher = "Addison-Wesley",
   address = "Reading, Massachusetts",
   edition = "Second",
   month = "10~" # jan,
   year = "\noopsort{1973b}1973",
   type = "Section",
   chapter = "1.2",
   pages = "10--119",
   note = "A full INBOOK entry",
}

@ARTICLE{Smith2005,
   author       = "J. S. Smith and G. W. Johnson", 
   journal      = "Philos. Trans. R. Soc. London, Ser. B", 
   title        = "", 
   year         = "2005", 

   volume       = "777", 
   pages        = "1395",
}

@UNPUBLISHED{Smith2010,
   author       = "W. J. Smith and T. J. Johnson and B. G. Miller", 
   title        = "Surface chemistry and preferential crystal orientation on a silicon surface", 
   note         = "{J. Appl. Phys.} (unpublished)", 
   
   month        = "", 
   year         = "",
}

@UNPUBLISHED{Smith2010a,
   author       = "V. K. Smith and K. Johnson and M. O. Klein", 
   title        = "Surface chemistry and preferential crystal orientation on a silicon surface", 
   note         = "{J. Appl. Phys.} (submitted)", 
   
   month        = "", 
   year         = "",
}

@UNPUBLISHED{unpublished-full,
   author = "Ulrich {\"{U}}nderwood and Ned {\~N}et and Paul {\={P}}ot",
   title = "Lower Bounds for Wishful Research Results",
   month = nov # ", " # dec,
   year = 1988,
   note = "Talk at Fanstord University (A full UNPUBLISHED entry)",
}

@MISC{JohnsonMillerSmith2007, 

   author       = "M. P. Johnson and K. L. Miller and K. Smith", 
   title        = "", 
   howpublished = "personal communication", 
   month        = "1~" # may, 
   year         = "2007", 
   note         = "",
}

@PROCEEDINGS{Smith2007, 
   title        = "AIP Conf. Proc.", 
   year         = "2007", 
   
   editor       = "J. Smith", 
   volume       = "841", 
   number       = "21", 
   series       = "", 
   address      = "", 
   month        = "", 
   organization = "", 
   publisher    = "", 
   note         = "", 
}

@PROCEEDINGS{proceedings-full,
   editor = "Wizard V. Oz and Mihalis Yannakakis",
   title = "Proc. Fifteenth Annual",
   number = 17,
   series = "All ACM Conferences",
   month = mar,
   year = 1983,
   address = "Boston",
   organization = "ACM",
   publisher = "Academic Press",
   note = "A full PROCEEDINGS entry",
}

@UNPUBLISHED{Burstyn2004,
   author       = "Y. Burstyn", 
   title        = "{Proceedings of the 5th International Molecular Beam Epitaxy Conference, Santa Fe, NM}", 
   note         = "(unpublished)", 
   
   month        = "5--8~" # oct, 
   year         = "2004",
}

@PROCEEDINGS{Quinn2001, 
   title        = "{Proceedings of the 2003 Particle  Accelerator Conference, Portland, OR, 12-16 May 2005}", 
   year         = "2001", 
   
   editor       = "B. Quinn", 
   address      = "New York", 
   publisher    = "Wiley", 
   note         = "Albeit the conference was held in 2005, it was the 2003 conference, and  the proceedings were published in 2001; go figure", 
}

@ARTICLE{Agarwal2001,
   author       = "A. G. Agarwal", 
   title        = "{Proceedings of the Fifth Low Temperature Conference, Madison, WI, 1999}", 
   journal      = "Semiconductors", 
   year         = "2001", 

   volume       = "66", 
   pages        = "1238", 
}

@ARTICLE{SmithDA01,
   author       = "R. Smith",
   title        = "Hummingbirds are our friends",
   journal      = {J. Appl. Phys. (these proceedings)},
   year         = "",
   volume       = "",
   number       = "",
   pages        = "",
   month        = "",
   note         = "Abstract No. DA-01",
}

@ARTICLE{Smith2007a, 
   author       = "J. Smith", 
   title        = "", 
   journal      = "Proc. SPIE", 
   year         = "2007", 

   volume       = "124", 
   pages        = "367", 
   note         = "Required title is missing", 
}

@TECHREPORT{techreport-full,
   author = "Tom T{\'{e}}rrific",
   title = "An {$O(n \log n / \! \log\log n)$} Sorting Algorithm",
   institution = "Fanstord University",
   type = "Wishful Research Result",
   number = "7",
   address = "Computer Science Department, Fanstord, California",
   month = oct,
   year = 1988,
   note = "A full TECHREPORT entry",
}

@TECHREPORT{Nelson1999, 
   author       = "J. Nelson", 
   type         = "{TWI Report}", 
   number       = "666/1999",
   institution  = "", 
   year         = jan # "~1999", 
   
   note         = "Required institution missing", 
}

@TECHREPORT{Fields2005, 
   author       = "W. K. Fields", 
   type         = "{ECE Report No.}", 
   number       = "AL944",
   institution  = "", 
   year         = "2005", 
   
   note         = "Required institution missing", 
}

@MISC{Zalkins2008, 

   author       = "Y. M. Zalkins", 
   title        = "", 
   howpublished = "e-print arXiv:cond-mat/040426", 
   month        = "", 
   year         = "2008", 
   note         = "",
}

@MISC{Nelson2005, 

   author       = "J. Nelson", 
   howpublished = "{U.S. Patent No.} 5,693,000", 
   year         = "12~" # dec # "~2005", 
}

@MASTERSTHESIS{Nelson1999a,
   author       = "J. K. Nelson", 
   title        = "", 
   school       = "New York University", 
   year         = "1999", 
   
   type         = "M.{S}. thesis", 
   address      = "", 
   month        = "", 
   note         = "", 
}

@MASTERSTHESIS{mastersthesis-full,
   author = "{\'{E}}douard Masterly",
   title = "Mastering Thesis Writing",
   school = "Stanford University",
   type = "Master's project",
   address = "English Department",
   month = jun # "-" # aug,
   year = 1988,
   note = "A full MASTERSTHESIS entry",
}

@PHDTHESIS{Smith2003,
   author       = "S. M. Smith", 
   title        = "", 
   school       = "Massachusetts Institute of  Technology", 
   year         = "2003", 
   
   type         = "{Ph.D.} thesis", 
   address      = "", 
   month        = "", 
   note         = "", 
}

@ARTICLE{KawaLin2003,
   author       = "S. R. Kawa and S.-J. Lin", 
   title        = "", 
   journal      = "J. Geophys. Res.", 
   year         = "2003", 

   volume       = "108", 
   number       = "D6", 
   pages        = "4201", 
   month        = "", 
   note         = "{DOI:10.1029/2002JD002268}", 
}

@PHDTHESIS{phdthesis-full,
   author = "F. Phidias Phony-Baloney",
   title = "Fighting Fire with Fire: Festooning {F}rench Phrases",
   school = "Fanstord University",
   type = "{PhD} Dissertation",
   address = "Department of French",
   month = jun # "-" # aug,
   year = 1988,
   note = "A full PHDTHESIS entry",
}

@BOOK{book-full,
   author = "Donald E. Knuth",
   title = "Seminumerical Algorithms",
   volume = 2,
   series = "The Art of Computer Programming",
   publisher = "Addison-Wesley",
   address = "Reading, Massachusetts",
   edition = "Second",
   month = "10~" # jan,
   year = "\noopsort{1973c}1981",
   note = "A full BOOK entry",
}

@BOOKLET{booklet-full,
   author = "Jill C. Knvth",
   title = "The Programming of Computer Art",
   howpublished = "Vernier Art Center",
   address = "Stanford, California",
   month = feb,
   year = 1988,
   note = "A full BOOKLET entry",
}

@INBOOK{ballagh2000,
   author    = "R. Ballagh and C.M. Savage",
   editor    = "C.M. Savage and M. Das",
   title     = "Bose-Einstein condensation: from atomic physics to quantum fluids, Proceedings of the 13th Physics Summer School",
   year      = "2000",
   publisher = "World Scientific",
   address   = "Singapore",
   eprint    = "cond-mat/0008070",
}

@inBook{Magnetism,
   author    = "W. Opechowski and R. Guccione",
   title     = "Introduction to the Theory of Normal Metals",
   volume    = "IIa",
   pages     = "105",
   editor    = "G. T. Rado and H. Suhl",
   booktitle = "Magnetism",
   publisher = "Academic Press",
   address   = "New York",
}

@INPROCEEDINGS{Magnetismb,
   author    = "W. Opechowski and R. Guccione",
   title     = "Introduction to the Theory of Normal Metals",
   editor    = "G. T. Rado and H. Suhl",
   booktitle = "Magnetism",
   volume    = "IIa",
   pages     = "105",
   publisher = "Academic Press",
   address   = "New York",
   year      = "1965",
}

@INBOOK{Smith80,
   author    = "J. M. Smith",
   title     = "Molecular Dynamics",
   publisher = "Academic",
   year      = "1980",
   address   = "New York",
   editor    = "C. Brown",
}

@article{ZS71,
   author       = "V. E. Zakharov and A. B. Shabat",
   year         = "1971",
   title        = "Exact theory of two-dimensional self-focusing and one-dimensional self-modulation of waves in nonlinear media",
   journal      = "Zh. Eksp. Teor. Fiz.",
   volume       = "61",
   pages        = "118--134",
   translation  = "Sov. Phys. JETP \textbf{34}, 62 (1972)"
}

@INCOLLECTION{Beutler1994a,
   author       = "E. Beutler", 
   year         = "1994", 
   booktitle    = "Williams Hematology", 
   edition      = "5", 
   chapter      = "7", 
   editor       = "E. Beutler and M. A. Lichtman and B. W. Coller and T. S. Kipps", 
   publisher    = "McGraw-Hill", 
   address      = "New York", 
   volume       = "2", 
   pages        = "654--662",
}

@INCOLLECTION{ballagh2000a,
   author       = "R. Ballagh and C.M. Savage",
   year         = "2000",
   title        = "Bose-Einstein condensation: from atomic physics to quantum fluids",
   booktitle    = "Proceedings of the 13th Physics Summer School",
   editor       = "C.M. Savage and M. Das",
   publisher    = "World Scientific",
   address      = "Singapore",
   eprint       = "cond-mat/0008070",
}

@INCOLLECTION{Magnetisma,
   author       = "W. Opechowski and R. Guccione",
   year         = "1965",
   title        = "Introduction to the Theory of Normal Metals",
   booktitle    = "Magnetism",
   editor       = "G. T. Rado and H. Suhl",
   publisher    = "Academic Press",
   address      = "New York",
   volume       = "IIa",
   pages        = "105",
}

@INCOLLECTION{Smith80a,
   author       = "J. M. Smith",
   year         = "1980",
   booktitle    = "Molecular Dynamics",
   editor       = "C. Brown",
   publisher    = "Academic",
   address      = "New York",
}

@INCOLLECTION{incollection-full,
   key          = "incol-ful",
   author       = "Daniel D. Lincoll",
   year         = 1977,
   title        = "Semigroups of Recurrences",
   booktitle    = "High Speed Computer and Algorithm Organization",
   edition      = "Third",
   series       = "Fast Computers",
   number       = 23,
   chapter      = 3,
   type         = "Part",
   editor       = "David J. Lipcoll and D. H. Lawrie and A. H. Sameh",
   publisher    = "Academic Press",
   address      = "New York",
   month        = sep,
   pages        = "179--183",
   note         = "A full INCOLLECTION entry",
}

@INPROCEEDINGS{inproceedings-full,
   author = "Alfred V. Oaho and Jeffrey D. Ullman and Mihalis Yannakakis",
   title = "On Notions of Information Transfer in {VLSI} Circuits",
   editor = "Wizard V. Oz and Mihalis Yannakakis",
   booktitle = "Proc. Fifteenth Annual ACM",
   number = 17,
   series = "All ACM Conferences",
   pages = "133--139",
   month = mar,
   year = 1983,
   address = "Boston",
   organization = "ACM",
   publisher = "Academic Press",
   note = "A full INPROCEDINGS entry",
}

@MANUAL{manual-full,
   author = "Larry Manmaker",
   title = "The Definitive Computer Manual",
   organization = "Chips-R-Us",
   address = "Silicon Valley",
   edition = "Silver",
   month = apr # "-" # may,
   year = 1986,
   note = "A full MANUAL entry",
}

@ARTICLE{caves1981,
   author       = "C. M. Caves",
   year         = "1981",
   journal      = "Phys.\ Rev.\ D",
   volume       = "23",
   pages        = "1693--1708",
}

@ARTICLE{aasi2013,
   author  = "J. Aasi and J. Abadie and B. P. Abbott and others",
   year    = "2013",
   journal = "Nat.\ Photon.",
   volume  = "7",
   pages   = "613--619",
}

@ARTICLE{giovannetti2011,
   author       = "V. Giovannetti and S. Lloyd and L. Maccone",
   year         = "2011",
   journal      = "Nat.\ Photon.",
   volume       = "5",
   pages        = "222--229",
}

@ARTICLE{giovannetti2004,
   author       = "V. Giovannetti and S. Lloyd and L. Maccone",
   year         = "2004",
   journal      = "Science",
   volume       = "306",
   pages        = "1330--1336",
}

@ARTICLE{resch2007,
   author       = "K. J. Resch and K. L. Pregnell and R. Prevedel and A. Gilchrist and G. J. Pryde and J. L. O'Brien and A. G. White",
   year         = "2007",
   journal      = "Phys.\ Rev.\ Lett.",
   volume       = "98",
   pages        = "223601",
}

@ARTICLE{slussarenko2017,
   author       = "S. Slussarenko and M. M. Weston and H. M. Chrzanowski and L. K. Shalm and V. B. Verma and S. W. Nam and G. J. Pryde",
   year         = "2017",
   journal      = "Nat.\ Photon.",
   volume       = "11",
   pages        = "700--703",
}

@INBOOK{barnett,
   author = "Stephen M. Barnett and Paul M. Radmore",
   title = "Methods in theoretical quantum optics",
   publisher = "Oxford University Press",
   year = "2002"
}

@INBOOK{vandenbos,
   author = "Adriaan Van den Boos",
   title = "Parameter estimation for scientist and engineers",
   publisher = "John Wiley and Sons",
   year = "2007",
}

@ARTICLE{dowling2008,
   author       = "J. P. Dowling",
   year         = "2008",
   journal      = "Contemp.\ Phys.",
   volume       = "49",
   pages        = "125--143",
}

@ARTICLE{holland1993,
   author       = "M. J. Holland and K. Burnett",
   year         = "1993",
   journal      = "Phys.\ Rev.\ Lett.",
   volume       = "71",
   pages        = "1355--1358",
}

@ARTICLE{dorner2009,
   author       = "U. Dorner and R. {Demkowicz-Dobrza{\'n}ski} and B. J. Smith and J. S. Lundeen and W. Wasilewski and K. Banaszek and I. A. Walmsley",
   year         = "2009",
   journal      = "Phys.\ Rev.\ Lett.",
   volume       = "102",
   pages        = "040403",
}

@ARTICLE{thekkadath2020,
   author       = "G. S. Thekkadath and M. E. Mycroft and B. A. Bell and C. G. Wade and A. Eckstein and D. S. Phillips and R. B. Patel and A. Buraczewski and A. E. Lita and T. Gerrits and S. W. Nam and M. Stobinska and A. I. Lvovsky and I. A. Walmsley",
   year         = "2020",
   journal      = "npj Quantum Inf.",
   volume       = "6",
   pages        = "89",
}

@ARTICLE{yurke1986,
   author       = "B. Yurke and S. L. McCall and J. R. Klauder",
   year         = "1986",
   journal      = "Phys.\ Rev.\ A",
   volume       = "33",
   pages        = "4033--4054",
}

@ARTICLE{hudelist2014,
   author       = "F. Hudelist and J. Kong and C. Liu and J. Jing and Z. Y. Ou and W. Zhang",
   year         = "2014",
   journal      = "Nat.\ Commun.",
   volume       = "5",
   pages        = "3049",
}

@ARTICLE{ou2012,
   author       = "Z. Y. Ou",
   year         = "2012",
   journal      = "Phys.\ Rev.\ A",
   volume       = "85",
   pages        = "023815",
}

@ARTICLE{qin2023,
   author       = "J. Qin and Y.-H. Deng and H.-S. Zhong and L.-C. Peng and H. Su and Y.-H. Luo and J.-M. Xu and D. Wu and S.-Q. Gong and H.-L. Liu and H. Wang and M.-C. Chen and L. Li and N.-L. Liu and C.-Y. Lu and J.-W. Pan",
   year         = "2023",
   journal      = "Phys.\ Rev.\ Lett.",
   volume       = "130",
   pages        = "070801",
}

@ARTICLE{marino2012,
   author       = "A. M. Marino and N. V. {Corzo Trejo} and P. D. Lett",
   year         = "2012",
   journal      = "Phys.\ Rev.\ A",
   volume       = "86",
   pages        = "023844",
}

@ARTICLE{frascella2021,
   author       = "G. Frascella and S. Agne and F. Ya. Khalili and M. V. Chekhova",
   year         = "2021",
   journal      = "npj Quantum Inf.",
   volume       = "7",
   pages        = "72",
}

@ARTICLE{chekhova2016,
   author       = "M. V. Chekhova and Z. Y. Ou",
   year         = "2016",
   journal      = "Adv.\ Opt.\ Photon.",
   volume       = "8",
   pages        = "104--155",
}

@ARTICLE{manceau2017,
   author       = "M. Manceau and G. Leuchs and F. Khalili and M. Chekhova",
   year         = "2017",
   journal      = "Phys.\ Rev.\ Lett.",
   volume       = "119",
   pages        = "223604",
}

@ARTICLE{giese2017,
   author       = "E. Giese and S. Lemieux and M. Manceau and R. Fickler and R. W. Boyd",
   year         = "2017",
   journal      = "Phys.\ Rev.\ A",
   volume       = "96",
   pages        = "053863",
}

@ARTICLE{lita2008,
   author       = "A. E. Lita and A. J. Miller and S. W. Nam",
   year         = "2008",
   journal      = "Opt.\ Express",
   volume       = "16",
   pages        = "3032--3040",
}

@ARTICLE{tse2019,
   author = "M. Tse and et al.",
   year   = "2019",
   journal = "Phys.\ Rev.\ Lett.",
   volume  = "123",
   pages   = "231107",
}

@ARTICLE{taylor2013,
   author = "M. A. Taylor and J. Janousek and V. Daria and 
             J. Knittel and B. Hage and H.-A. Bachor and W. P. Bowen",
   year   = "2013",
   journal = "Nat.\ Photon.",
   volume  = "7",
   pages   = "229--233",
}

@ARTICLE{taylor2016,
   author = "M. A. Taylor and W. P. Bowen",
   year   = "2016",
   journal = "Phys.\ Rep.",
   volume  = "615",
   pages   = "1--59",
}

@ARTICLE{demkowicz2015,
   author = "R. {Demkowicz-Dobrza{\'n}ski} and M. Jarzyna and 
             J. Ko{\l}ody{\'n}ski",
   year   = "2015",
   journal = "Prog.\ Opt.",
   volume  = "60",
   pages   = "345--435",
}

@ARTICLE{laissue2017,
   author       = "P. P. Laissue and R. A. Alghamdi and P. Tomancak and E. G. Reynaud and H. Shroff",
   year         = "2017",
   journal      = "Nat.\ Methods",
   volume       = "14",
   pages        = "657--661",
}

@ARTICLE{ligo2011,
   author       = "{The LIGO Scientific Collaboration}",
   year         = "2011",
   journal      = "Nat.\ Phys.",
   volume       = "7",
   pages        = "962--965",
}

@ARTICLE{you2021,
   author       = "C. You and M. Hong and P. Bierhorst and A. E. Lita and S. Glancy and S. Kolthammer and E. Knill and S. W. Nam and R. P. Mirin and O. S. Maga{\~n}a-Loaiza and T. Gerrits",
   year         = "2021",
   journal      = "Appl.\ Phys.\ Rev.",
   volume       = "8",
   pages        = "041406",
}

@ARTICLE{heilmann2016,
   author       = "R. Heilmann and J. Sperling and A. Perez-Leija and M. Gr{\"a}fe and M. Heinrich and S. Nolte and W. Vogel and A. Szameit",
   year         = "2016",
   journal      = "Sci.\ Rep.",
   volume       = "6",
   pages        = "19489",
}

@ARTICLE{wildfeuer2009,
   author       = "C. F. Wildfeuer and A. J. Pearlman and J. Chen and J. Fan and A. Migdall and J. P. Dowling",
   year         = "2009",
   journal      = "Phys.\ Rev.\ A",
   volume       = "80",
   pages        = "043822",
}

@ARTICLE{glauber1967a,
   author  = "B. R. Mollow and R. J. Glauber",
   year    = "1967",
   journal = "Phys.\ Rev.",
   volume  = "160",
   pages   = "1076--1096",
}

@ARTICLE{glauber1967b,
   author  = "B. R. Mollow and R. J. Glauber",
   year    = "1967",
   journal = "Phys.\ Rev.",
   volume  = "160",
   pages   = "1097--1108",
}

@BOOK{gradshteyn2007,
   author    = "I. S. Gradshteyn and I. M. Ryzhik",
   title     = "Table of Integrals, Series, and Products",
   edition   = "7th",
   publisher = "Academic Press",
   year      = "2007",
}

@ARTICLE{li2025,
   author  = "P. Z. Li and W. Zhang and Z. F. Feng and Z. Wang and Q. X. Ma and J. Q. Zhong and W. Miao and Q. J. Yao and J. Li and S. C. Shi",
   year    = "2025",
   journal = "APL Photon.",
   volume  = "10",
   pages   = "120802",
}

@BOOK{haus2000,
   author    = "H. A. Haus",
   title     = "Electromagnetic Noise and Quantum Optical Measurements",
   publisher = "Springer-Verlag, Berlin",
   year      = "2000",
}

@ARTICLE{boyd2008,
   author  = "R. W. Boyd and G. S. Agarwal and K. W. C. Chan and A. K. Jha and M. N. Sullivan",
   year    = "2008",
   journal = "Opt. Commun.",
   volume  = "281",
   pages   = "3732",
}

@ARTICLE{lopaeva2013,
   author  = "E. D. Lopaeva and I. Ruo Berchera and I. P. Degiovanni and S. Olivares and G. Brida and M. Genovese",
   year    = "2013",
   journal = "Phys.\ Rev.\ Lett.",
   volume  = "110",
   pages   = "153603",
}

@ARTICLE{gerrits2011,
  author  = "T. Gerrits and M. J. Thomas and B. Calkins and A. E. Lita and W. H. P. Pernice and H. X. Tang and R. P. Mirin and S. W. Nam",
  year    = "2011",
  journal = "Opt.\ Express",
  volume  = "19",
  pages   = "24434--24447",
}

@ARTICLE{escale2020,
  author  = "M. Reig Escal{\'e} and D. Pohl and A. Flry and M. Timofeeva and R. Grange",
  year    = "2020",
  journal = "Opt.\ Express",
  volume  = "28",
  pages   = "26822--26832",
}

@ARTICLE{zhang2017,
  author  = "M. Zhang and C. Wang and R. Cheng and A. Amir and S. Shayan and M. Lon{\v{c}}ar",
  year    = "2017",
  journal = "Optica",
  volume  = "4",
  pages   = "1536--1537",
}

@ARTICLE{kranias2025,
  author  = "J. Kranias and G. Thekkadath and K. Heshami and A. Z. Goldberg",
  year    = "2025",
  journal = "Quantum",
  volume  = "9",
  pages   = "1619",
}

@ARTICLE{giese2026,
  author  = "C. Oglialoro and G. J. Machado and F. Farsch and D. F. Urrego and A. A. Padilla and R. B. Patel and I. A. Walmsley and M. Gr\''{a}fe and J. P. Torres and E. Giese",
  year    = "2026",
  journal = "	arXiv:2601.04139 [quant-ph]",
}

@ARTICLE{chekhova2006,
  author  = "O. A. Ivanova and T. Sh. Iskhakov and A. N. Penin and M. V. Chekhova",
  year    = "2006",
  journal = "Quantum Electron.",
  volume  = "36",
  pages   = "951–-956",
}

@ARTICLE{spasibko2012,
  author  = "K. Yu. Spasibko and T. Sh. Iskhakov and M. V. Chekhova",
  year    = "2012",
  journal = "Opt. Express",
  volume  = "20",
  pages   = "7507--7515",
}

@ARTICLE{thomaspeter2011,
  author  = "N. Thomas-Peter and B. J. Smith and A. Datta and L. Zhang and U. Dorner and I. A. Walmsley",
  year    = "2011",
  journal = "Phys. Rev. Lett.",
  volume  = "107",
  pages   = "113603",
}

@ARTICLE{zhenghao2025,
   author  = "Z. Li and M. J. H. Kendall and G. J. Machado and 
               R. Zhu and E. Mer and H. Zhan and A. Zhang and 
               S. Yu and I. A. Walmsley and R. B. Patel",
   year    = "2025",
   journal = "Optica Quantum",
   volume  = "3",
   pages   = "246--255",
}

@ARTICLE{levine2012,
	author = "Z. H. Levine and T. Gerrits and A. L. Migdall and D. V. Samarov and B. Calkins and A. E. Lita and S. W. Nam",
    year = "2012",
	journal = "Journal of the Opt. Soc. of Am. B",
	volume = "29",
	pages = "2066",
}

@ARTICLE{demkowicz2013,
    author = "R. {Demkowicz-Dobrza{\'n}ski} and K. Banaszek and R. Schnabel",
    year = "2013",
    journal = "Phys. Rev. A",
    volume = "88",
    pages = "041802(R)",
}

\end{document}